\documentclass[twocolumn,showpacs,preprintnumbers,amsmath,amssymb]{revtex4}
\usepackage{graphicx}% Include figure files
\usepackage{dcolumn}% Align table columns on decimal point
\usepackage{bm}% bold math

\usepackage{color}

\begin{document}

\title{Curvature-induced radiation of\\surface plasmon polaritons propagating around bends}% Force line breaks with \\

\author{Keisuke Hasegawa}
\author{Jens U. N\"ockel}
\author{Miriam Deutsch}
 \email{miriamd@uoregon.edu}
\affiliation{Department of Physics, University of Oregon, 1371 E 13th Ave., Eugene, OR, USA 97403 }

\date{\today}% It is always \today, today,

\begin{abstract}
We present a theoretical study of the curvature-induced radiation of surface plasmon polaritons propagating around bends at metal-dielectric interfaces. We explain qualitatively how the curvature leads to distortion of the phase front, causing the fields to radiate energy away from the metal-dielectric interface. We then quantify, both analytically and numerically, radiation losses and energy transmission efficiencies of SPPs propagating around bends with varying radii- as well as sign-of-curvature.
\end{abstract}

\maketitle

\section{\normalsize Introduction}
\label{sec: intro}

Surface plasmon polaritons (SPPs), coupled modes of plasmons and photons, are low-dimensional excitations propagating at metal-dielectric interfaces. As such, SPPs are confined to lateral dimensions of the order $\lambda/10$, with $\lambda$ the vacuum wavelength of light, enabling plasmonic devices which are more compact than existing photonic equivalents. Owing to their large electromagnetic (EM) field intensity at the interfaces, SPPs are highly sensitive to the surface morphology, thus allowing the realization of metallic structures capable of controlling and manipulating light on the nano-scale. This, in addition to advances in nanofabrication technologies, has led to a growing interest in realizing ultra-compact plasmon-based integrated circuits. Recent studies of plasmonic manipulation include SPP waveguiding and bending in patterned metallic films\cite{Bozhevolyni01-1,Bozhevolyni01-2}, guiding via resonant energy transfer in ordered arrays of metal nanoparticles\cite{Quinten98}, as well as demonstrations of SPP prisms, lenses\cite{Hohenau05}, mirrors, beamsplitters, and interferometers\cite{Ditlbacher02}.

As the need for integration of compact lightwave devices is growing, it is necessary that we obtain an understanding of the fundamental properties of SPP propagation and manipulation in these environments. Of particular interest is the development a quantitative theory of curvature-induced radiative energy loss in SPPs propagating at curved metal-dielectric interfaces. This enables the determination of SPP propagation efficiencies when the radius of curvature is smaller than or comparable to the signal wavelength\cite{Hasegawa04, Grischkowsky, Smith}. Careful analysis of the relations between the propagation efficiency and the interface curvature is essential when designing plasmonic devices, as it should set a limit on the radius of curvature and, subsequently, on feature size in plasmonic-circuits. Several recent studies have addressed surface plasmon waveguiding around bends. However, these studies are primarily focused on \emph{in-plane} guiding\cite{Weeber05, Steinberger06, Berini06, Radko06} or on using multiple-interface geometries such as metal-dielectric-metal waveguides\cite{Veronis05} or long-range SPP waveguides\cite{Kim06}. To date, SPP guiding on a single bent interface remains largely unknown.

This article presents a theoretical analysis of curvature-induced radiation emitted by SPPs propagating around bends at metal-dielectric interfaces. In Sec.~\ref{sec:geomtery} we introduce the geometry of our system, followed by a simple qualitative analysis in Sec.~\ref{sec:qualitative} describing the radiative nature of SPP propagation around bends. This enables us to develop a more rigorous analytical approach for calculating losses and propagation efficiencies, which we present in Sec.~\ref{sec:analytical}. Section~\ref{sec:numerical} provides a subsequent numerical analysis, and its results are compared with the analytical results. A novel resonator-based method for enhancing the transmittance is presented in Sec.~\ref{sec:resonator}. Conclusions are presented in Sec.~\ref{sec:conclusions}.

   \begin{figure}[h]
   \includegraphics[width=8cm]{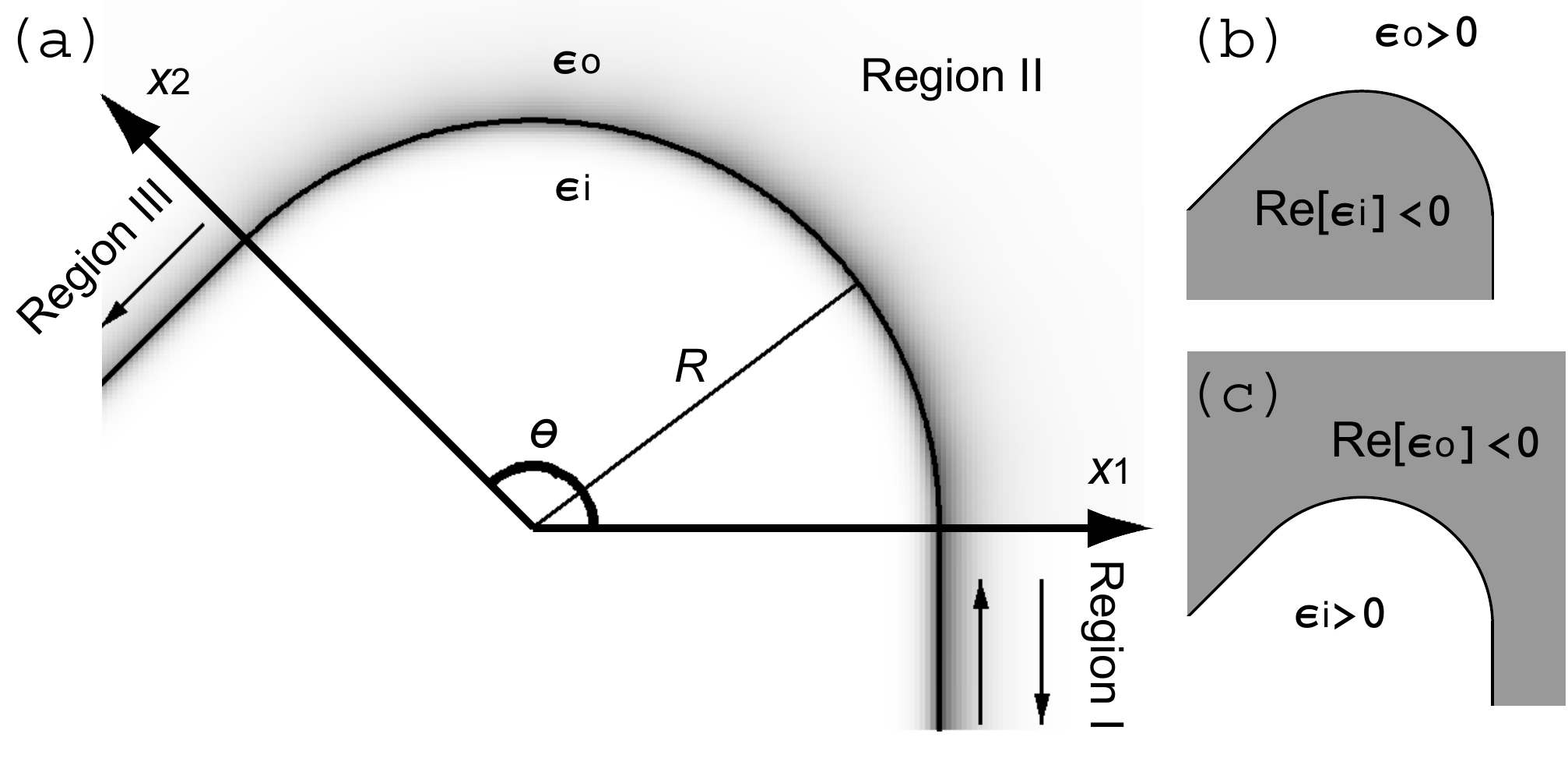}
   \caption{ \label{Fig1}
(a) The geometry of the curved metal-dielectric interface. A rounded edge, characterized by a permittivity $\epsilon_i$, has a bend angle $\theta$ and a finite bend radius $R$. The bend is confined to the region of space shown, with center of curvature at the origin. The rest of space is occupied by a medium with $\epsilon_o$. Axes $x_1$ and $x_2$ extend along the boundaries between regions I and II and regions II and III, respectively. In the $x_1-x_2$ plane, regions I and III are semi-infinite. The system is also infinite in extent along the entire $z$ axis. Arrows indicate incident and reflected fields in region I, and transmitted field in region III. In addition, (a) shows the intensity of SPPs in greyscale. The intensity distribution is calculated for SPPs traveling around a metallic corner (b), using the single-mode approximation developed in the text. We also consider the complementary configuration, in which SPPs propagate around a dielectric void (c).}
   \end{figure}

\section{\normalsize Geometry of the curved metal-dielectric interface}
\label{sec:geomtery}

The geometry of our study is shown in Fig.~\ref{Fig1}(a). The system consists of a curved metal-dielectric interface occupying region II in space, matching smoothly to planar and semi-infinite interfaces in regions I and III. The axes $x_1$ and $x_2$ define the boundaries between the three regions, and the system is infinite in the $z$ direction. The rounded edge is characterized by a permittivity $\epsilon_i$, with a fixed radius $R$ and a finite bend angle $\theta$. The surrounding space is characterized by a permittivity $\epsilon_o$. Propagating SPPs of frequency $\omega$ are incident from region I onto the boundary at $x_1$, and their counterclockwise transmission through region II into region III is analyzed. We initially consider the case of SPPs propagation around a \emph{metallic corner} with $\hbox{Re}[\epsilon_i]<0$ and $\epsilon_o>0$, depicted schematically in Fig.~\ref{Fig1}(b). This is essentially a segment of an infinitely long metal cylinder. We then analyze the complementary reverse geometry shown in Fig.~\ref{Fig1}(c), in which SPPs propagate around an open \emph{dielectric void}.

\section{\normalsize A Qualitative analysis of bend-induced radiation}
\label{sec:qualitative}

The generalized dispersion relation of SPPs propagating at a metal-dielectric (i.e. \emph{anisotropic}) interface is
\begin{equation}
\epsilon_{o,i}\frac{\omega^2}{c^2}=k_{\parallel}^2+k_{\perp o,i}^2
\end{equation}
where $k_{\parallel}$ and $k_{\perp o,i}$ are the components of the $k$-vector parallel and perpendicular to the interface, respectively. For a surface-guided mode, we require $k_{\parallel}> \sqrt{\epsilon_o}\omega/c$ such that $k_{\perp o,i}$ becomes imaginary, hence non-radiative. Rewriting this simple expression yields $\omega/k_{\parallel}<c\sqrt{\epsilon_o}$. Simply stated, the phase velocity in the direction parallel to the interface, $v_{\parallel}\equiv \omega/k_{\parallel}$, cannot exceed the speed of light, $c/\sqrt{\epsilon_o}$, in order to sustain non-radiative guiding.

When propagating around a bend, $v_{\parallel}$ acquires a radial dependence, with EM fields more distant from the interface travelling at greater phase velocities. Thus, there exists a threshold radius, $r^{\star}$, where the parallel phase velocity reaches the speed of light. Beyond $r^{\star}$ the EM field of the SPP becomes radiative. As a result, SPPs can be guided along curved interfaces with negligible radiation loss as long as the fields are confined near the metal-dielectric interface and do not extend beyond $r^{\star}$. This claim is verified analytically in the next section.

\section{\normalsize Transmission and reflection at bent interfaces}
\label{sec:analytical}

In order to quantify the degree of radiation loss and the propagation efficiency of SPPs around the bend we now exploit known solutions of Maxwell's equations describing angular propagation of EM waves at the surface of an infinitely long metal cylinder. These solutions are applicable in region II. The magnetic field in the dielectric is hence given by
\begin{equation}{\bf B}=\hat{\bf z}\sum_{\{n\}}\Bigl[A_n^+e^{+in\phi}+A_n^-e^{-in\phi}
\Bigr]H^{(1)}_n(k_or)e^{-i\omega t},\label{regionII}
\end{equation}
where $k_o=\omega\sqrt{\epsilon_o}/c$ and $H^{(1)}_n$ is the Hankel function of the first kind. The set of mode indices $\{n\}$, denoting radial excitations, is determined by the metal boundary matching equation
\begin{equation}0=\frac{1}{k_i}\frac{J'_n(k_iR)}{J_n(k_iR)}-
\frac{1}{k_o}\frac{H^{(1)'}_n(k_oR)}{H^{(1)}_n(k_oR)},\label{findn}
\end{equation}
where $k_i=\omega\sqrt{\epsilon_i}/c$, $J_n$ is the Bessel function, and the prime denotes differentiation with respect to the argument. Since region II comprises only a segment of a full cylinder (i.e. $\theta<2\pi$,) periodic boundary conditions need not be satisfied, and $n$ is therefore not constrained to integer values. In fact, since we choose the frequency $\omega$ to be real-valued, one finds $n$ to have a non-vanishing imaginary part as well. The latter is a consequence of radiation loss and absorption in the bend.

The azimuthal dependence of the SPP wave is given by the standard expression $\exp[\pm in\phi]$, where $\phi$ is the measured from the $x_1$ axis. Thus only solutions with $\hbox{Im}[n]\geq 0$ are admissible, describing attenuated propagation. Away from the interface in the dielectric region, where $k_o r\gg \hbox{Re}[n]$ is satisfied, we may use the asymptotic form of the Hankel function: $H^{(1)}_n(k_or)\approx\sqrt{2/\pi k_or}\exp[i(k_or-(2n+1)\pi/4)]$. Each mode can now be written as \begin{equation}{\bf B}_n\sim e^{\pm in\phi} \frac{e^{i(k_or-\omega t)}}{\sqrt{r}}\label{basymptotic}\end{equation} thus recovering the expected free propagating cylindrical wave form.

In general, Eq.~(\ref{findn}) cannot be solved algebraically, and it is necessary to employ a numerical method to find the set $\{n\}$. One convenient numerical approach employs graphical plotting of the right-hand side of Eq.~ (\ref{findn}) as functions of both $\hbox{Re}[n]$ and $\hbox{Im}[n]$, and identifying its zeros in the complex $n$-plane.

\begin{figure}[t]
   \includegraphics[width=7cm]{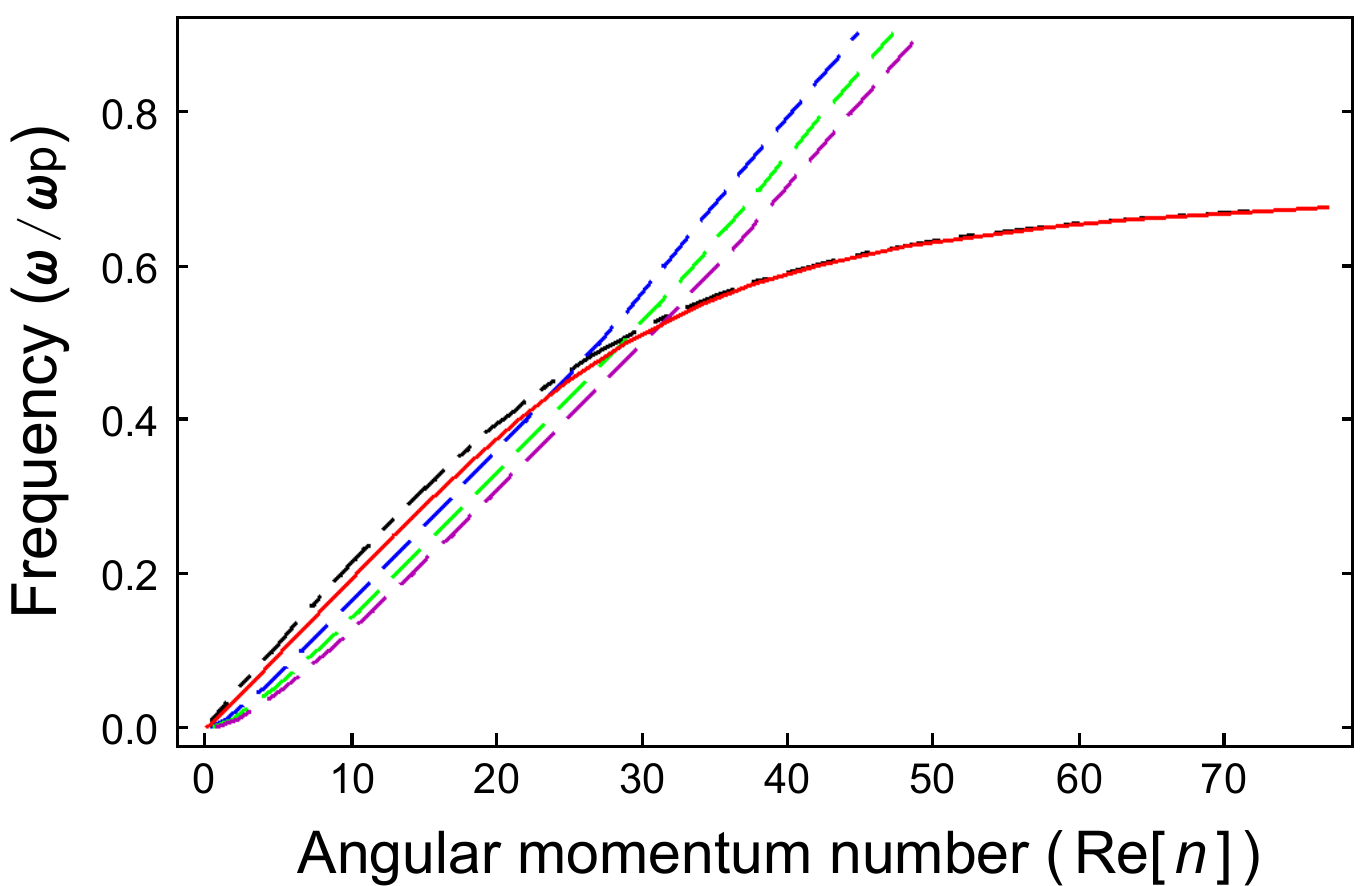}
   \caption{ \label{Fig2}
Dispersion relations of various transverse-magnetic modes on a metal cylinder with $R=2\mu\hbox{m}$. The dispersion relation for the fundamental mode (solid line) closely follows that of SPPs on a flat metal-dielectric interface (dash-dotted line). The horizontal scale of the latter is normalized by the factor $R$, so that the curve is a plot of $\omega$ as function of $kR$, instead of $k$. Dispersion relations of modes with higher radial excitations are also shown (dashed lines). These modes are less confined to the interface and do not approach $\omega_p/\sqrt{2}$ asymptotically. Higher radial excitation modes with even weaker surface confinement exist (not shown), all lying to the right of the three dashed lines shown.}
   \end{figure}

In principle, the set of solutions denoted by $\{n\}$ is infinite. In practice, we find that a single mode of this set dominates the propagation problem we are analyzing in this work. The set $\{n\}$ contains a \emph{fundamental mode} (i.e. the surface plasmon mode), which we label with mode index $m\in \{n\}$. The EM field of this mode is concentrated near the metal-dielectric interface. The solid line in Fig.~\ref{Fig2} shows the dispersion relation of this fundamental mode. In the limit of large momentum, it asymptotically approaches $\omega_{sp}\equiv\omega_p/\sqrt{1+\epsilon_o}$, the conventional limit for surface plasmon modes. Also shown in Fig.~\ref{Fig2} are the dispersion relations of modes in the set $\{n\}$ with higher radial excitation values (dashed lines). We see that these modes do not approach $\omega_{sp}$ asymptotically. Compared to the fundamental, these modes are also found to be less confined to the interface. As the radial excitation number increases even further, the modes corresponding to these values of $n$ all lie to the right of the three non-fundamental modes shown in Fig.~\ref{Fig2}. As expected from the fundamental mode, $\hbox{Im}[m]$ is the smallest of the set $\{n\}$ since the confinement of this mode to the interface is maximal, hence its radiation losses are the lowest. This is generally true for metals such as gold and silver, where the absorption losses are relatively low\cite{Footnote1}.

As shown in Fig.~\ref{Fig2}, away from $\omega_{sp}$ the fundamental mode has the smallest angular momentum ($\approx kR$) at a given frequency. To understand this, we consider the following: The angular momentum is ${\bf L}=\int ({\bf r}\times{\bf S}/c^2)dr^3$, where $\bf{S}$ is the Poynting vector. Compared to other modes, the fundamental mode is more strongly confined near the surface. Since $\bf{S}$ for this mode is significant only at $r\simeq R$, the integrand is minimal (as long as $\bf{S}$ of the fundamental mode is not disproportionately large.) However, as the frequency approaches $\omega_{sp}$, $\bf{S}$ increases asymptotically, which overcompensates for the smaller value of $\bf{r}$. This is seen in Fig.~\ref{Fig2} as a crossover of the fundamental mode and the other depicted modes, such that at $\omega\sim\omega_{sp}$ its angular momentum is higher than that of the less confined modes.

The radiative nature of the SPP solutions described by Eq. (\ref{regionII}) is also consistent with the qualitative analysis presented in the previous section. This can be verified by analyzing the nature of the Hankel function $H^{(1)}_n$. The Hankel function is non-oscillatory (depicting non-radiative fields) while its argument is smaller than the order $\hbox{Re}[n]$. When the argument exceeds $\hbox{Re}[n]$, $H^{(1)}_n$ starts to approach its oscillatory (radiating) form, as shown by the asymptotic expression leading to Eq. (\ref{basymptotic}). The non-oscillatory--to--oscillatory transition occurs when the argument and the order are approximately equal. We are thus led to conclude that each mode of Eq. (\ref{regionII}) undergoes a non-radiative--to--radiative transition at a radius given by $k_o r\approx\hbox{Re}[n]$.

The transition radius described above coincides with the previously introduced threshold radius, $r^{\star}$. By definition, at $r=r_n^{\star}$ we require $v_{\parallel}=c/\sqrt{\epsilon_o}$, where the mode index {n} is added to  $r^{\star}$ since each mode with index $n$ has a different threshold radius $r_n^{\star}$ in region II. The phase flow in the parallel direction is characterized by $\exp(i\hbox{Re}[n]\phi)$, and its associated wavenumber and phase velocity are $k_{\parallel}=\hbox{Re}[n]/r$ and $v_{\parallel}=\omega/k_{\parallel}=\omega r/\hbox{Re}[n]$, respectively. We thus obtain
\begin{equation}
r_n^{\star}=\frac{\hbox{Re}[n]}{k_o},\label{rstar}
\end{equation}
verifying that the threshold radius is indeed the non-radiative--to--radiating-SPP transition radius. From this it follows that the fundamental mode $m$, with lowest angular momentum ( for $\omega<\omega_{sp}$,) also has the smallest threshold radius for radiation:
\begin{equation}
r_m^{\star}=\frac{\hbox{Re}[m]}{k_o}\approx \hbox{Re}\left[\sqrt{\frac{\epsilon_i}{\epsilon_i+\epsilon_o}}\right]R.\label{rstarm}
\end{equation}
The approximation above follows from $\hbox{Re}[m]\approx kR$, where $k=\omega\sqrt{\epsilon_i\epsilon_o/(\epsilon_i+\epsilon_o)}/c$ is the wavenumber of SPPs propagating at a flat metal-dielectric interface. The validity of this approximation is confirmed in Fig.~\ref{Fig2}, where it is shown that the dispersion relation of the fundamental mode is very close to that of SPPs at a flat interface.

\begin{figure}[t]
   \includegraphics[width=8cm]{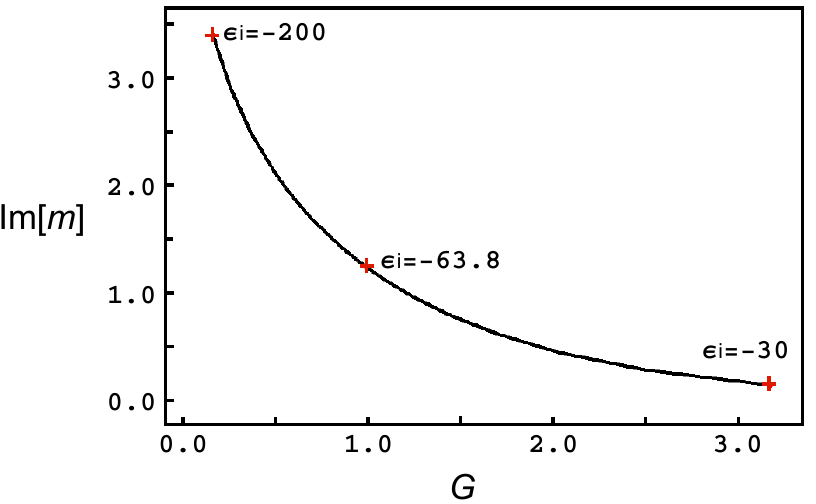}
   \caption{ \label{Fig3}
   Plot of $\hbox{Im}[m]$ as function of $G$ for $\omega R/c=1000$ and $\epsilon_o=1$. The premittivity $\epsilon_i$ is real and varies from $\epsilon_i=-30$ to $-200$.
}
   \end{figure}

The above analysis implies that when the majority of the SPP field is condfined to a region with $r<r^{\star}_m$, we should expect a negligible radiation loss during propagation. On the dielectric side of the interface the SPP field decays as $\exp\{-\hbox{Re}[\gamma_o(r-R)]\}$, with the decay coefficient is given by $\gamma_o=\omega\epsilon_o\sqrt{-1/(\epsilon_i+\epsilon_0)}/c$. We see that the field confinement in the dielectric is therefore characterized by $\hbox{Re}[\gamma_o]$.
%The field decay coefficient is given by $\gamma_o=\omega\epsilon_o\sqrt{-1/(\epsilon_i+\epsilon_0)}/c$. The SPP field confinement in the dielectric medium is therefore characterized by $\hbox{Re}[\gamma_o]$, and it decays away from the interface and into the dielectric as $\exp\{-\hbox{Re}[\gamma_o(r-R)]\}$,
Thus, we conclude that the radiation loss is insignificant if $G>1$, where $G$ is the exponent evaluated at $r=r^{\star}_m$:
\begin{equation}
G\left(\epsilon_i,\epsilon_o,\omega R/c\ \right)\equiv \hbox{Re}[\gamma_o](r^{\star}_m-R).\label{Gfactor}
\end{equation}
In Fig.~\ref{Fig3} we plot $\hbox{Im}[m]$ as function of $G$. The metal is assumed lossless (i.e. $\hbox{Im}[\epsilon_i]=0$), therefore $\hbox{Im}[m]$ is related solely to radiative loss. We find that $\hbox{Im}[m]$ is a monotonically decreasing function of $G$, as expected from the arguments above. Moreover, when $G=1$ $\hbox{Im}[m]=1.24\sim1$, confirming again that $G=1$ is a very good approximation of the threshold where SPP propagation changes from significantly radiative ($\hbox{Im}[m]\gtrsim 1$) to mostly non-radiative ($\hbox{Im}[m]\lesssim 1$).

\bigskip

To quantify the radiation loss and propagation efficiency around the bend it is now necessary to consider the coupling of these metal-cylinder modes to planar-interface SPP modes (i.e. propagating modes in regions I and III). The magnetic field of the SPP incident from Region I is given by
\begin{equation} {\bf B}=\hat{\bf z}A \exp\bigl(iky_1-i\omega t\bigr)\left\{\begin{array}{r@{\quad \quad}l}
\exp\bigl[-\gamma_i (x_1-R)\bigr] & x_1\geq 0 \\ \exp\bigl[+\gamma_o (x_1-R)\bigr]& x_1<0
\end{array}\right.\end{equation}
with $A$ the amplitude and $\gamma_i=-\omega\epsilon_i\sqrt{-1/(\epsilon_i+\epsilon_0)}/c$. Similar expressions hold for the SPPs reflected into region I and the fields transmitted into region III.

Calculating exact values of the transmission and reflection coefficients is often impractical, since it requires matching an infinite number of modes in Eq. (\ref{regionII}) to planar SPP modes in regions I and III. The mode matching must be carried out along the entire spatial extent of the $x_1$ and $x_2$ axes at the boundaries. However, in certain cases it is possible to derive relatively simple approximate expressions for the required field coefficients. This is because although each mode of region II has an EM field profile normal to the surface which does not exactly match the field profile of the incident planar SPP, \emph{the fundamental mode of region II minimizes this spatial mismatch}. Non-fundamental modes are characterized by EM fields which are less confined to the surface, therefore their field profiles deviate more strongly from those of the planar SPPs whose fields are always strongly bound at the metal-dielectric interface.

\begin{figure}[t]
   \includegraphics[width=5cm]{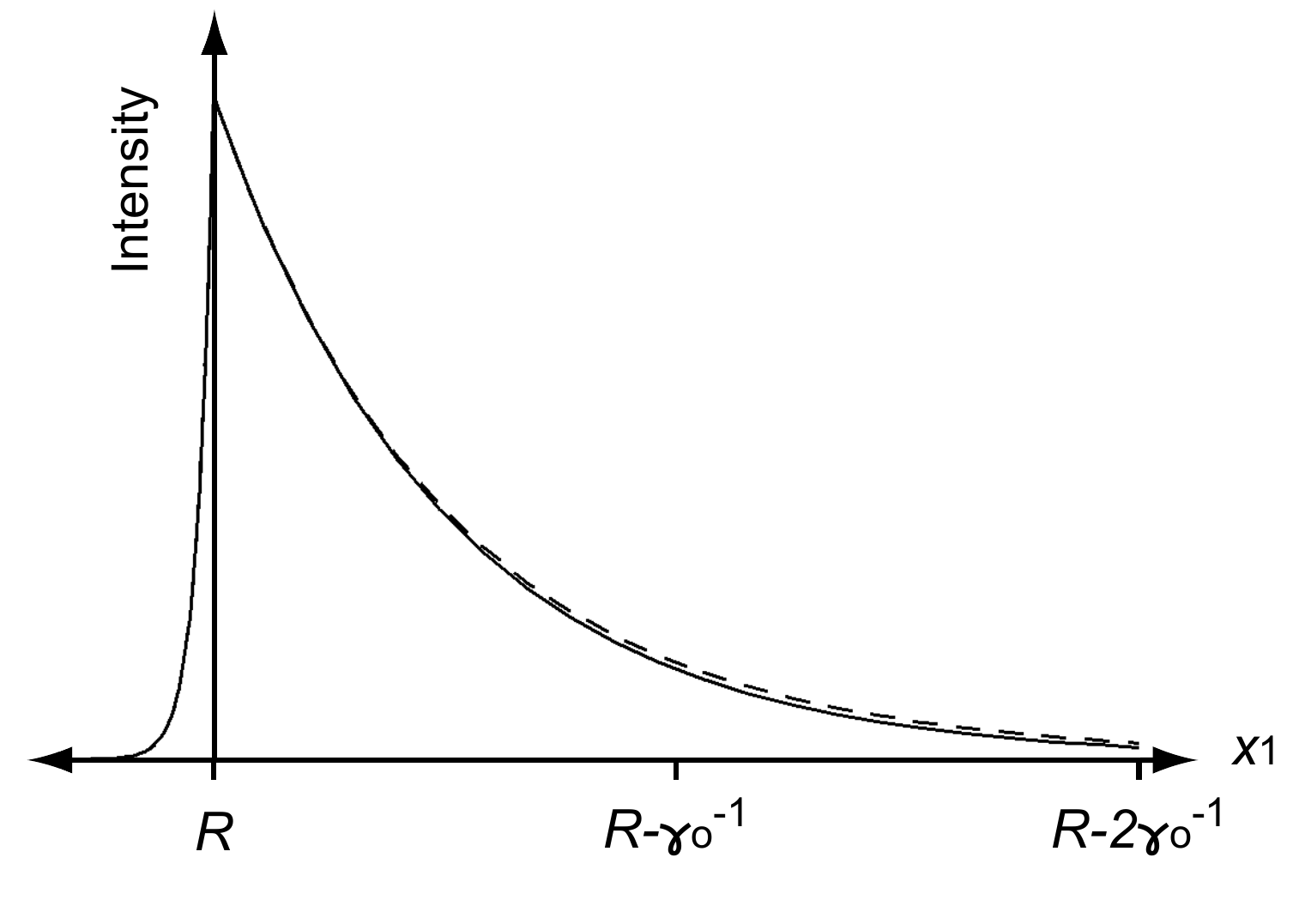}
   \caption{ \label{Fig4}
Comparison of intensity profiles of SPPs in region I (solid
line) and the fundamental mode in region II (dashed line),
calculated for $\omega R/c=800$. The profile mismatch is barely
visible, indicating that SPPs in region I couple predominately
to the fundamental mode in region II. }
   \end{figure}

In the short wavelength limit, it is possible to show that incident SPP and the fundamental mode have \emph{identical} field profiles near the interface. In this limit, the field distribution of the fundamental mode in the radial direction can be obtained by taking the appropriate limit of the Bessel equation $r^2d^2f/dr^2+rdf/dr+(k_{o,i}^2r^2-m^2)f=0$ to which $J_m(k_ir)$ and $H_m^{(1)}(k_or)$ are solutions, respectively. As $\omega R/c\rightarrow \infty$, the curvature of the metal surface becomes insignificant and $m$ approaches $kR$. With $r=R+x$, in the limit $x\ll R$ the Bessel equation reduces to $R^2d^2f/dx^2+(k_{i,o}^2-k^2)R^2f=0$, to which the solutions are exponentials. Thus, noting that $\gamma_{i,o}^2=k^2-k_{i,o}^2$, $J_m(k_ir)\sim \exp(\gamma_ir)$ and $H_m^{(1)}(k_or)\sim \exp(-\gamma_o r)$ which is identical to the behavior of the SPP fields in regions I and III near the interface. Far from the interface where the condition $x\ll R$ no longer holds, $J_m(k_ir)$ and $H_m^{(1)}(k_or)$ no longer exhibit exponentially decaying behaviors. However, due to the exponential decay near the interface, their values are small far from the interface, making the mismatch negligible along the entire $x_1$ and $x_2$ axes. This is illustrated in Fig.~\ref{Fig4}, where we show that the intensity profiles of the fundamental mode and the planar SPP are well matched in the short wavelength limit.

We are thus led to conclude that in the short wavelength limit planar SPPs couple predominately to the fundamental mode, and neglect their coupling to all other non-fundamental modes. For this reason, under this single-mode approximation, it is necessary to consider only a small number of modes: the incident and reflected SPPs in region I, the clockwise and counterclockwise propagating fundamental mode in region II, and the transmitted SPP in region III. These modes are matched at a single point on at each axis, at a distance R from the origin, via the standard Maxwell boundary conditions. Our analysis above ensures that the boundary conditions are then approximately satisfied over the entire extent of the axes. The problem of quantifying the propagation efficiency has now essentially become one dimensional (1D), and it is mathematically analogous to scattering from a 1D finite potential well\cite{Mekis96}. However, since the allowed $m$ values are always complex, bound-state solutions in this type of well do not exist. This distinguishes SPPs at curved surfaces from waveguide bends enclosed on all sides by infinite potential walls\cite{Sols90}.

Applying the appropriate boundary conditions to the fields at the $x_1$ and $x_2$ boundaries results in expressions for the transmittance $\textsf{T}$ and reflectance $\textsf{R}$:
\begin{equation}\textsf{T}=\left|\frac{4mkR}{-e^{im\theta}(m-kR)^2+
e^{-im\theta}(m+kR)^2}\right|^2\label{transmittance}
\end{equation}
\begin{equation}\textsf{R}=\left|\frac{2\sin(m\theta)(m^2-k^2R^2)}
{-e^{im\theta}(m-kR)^2+e^{-im\theta}(m+kR)^2}\right|^2.\label{reflectance}
\end{equation}
In the presence of significant absorption or radiation loss, such that $\hbox{Im}[m]\theta\gg 1$, these expressions become
\begin{equation}\textsf{T}\approx
16\frac{|mkR|^2}{|m+kR|^4}e^{-2\hbox{\scriptsize Im}[m]\theta}
\end{equation}
\begin{equation}\textsf{R}\approx\left|\frac{m-kR}
{m+kR}\right|^2.\label{reflectance2}
\end{equation}
As $\omega R/c\rightarrow\infty$, these expressions become exact.

In general, conservation of energy is expressed by $1=\textsf{T}+\textsf{R}+\textsf{P}+\textsf{A},$ where $\textsf{P}$ and $\textsf{A}$ are the radiation and absorption loss coefficient, respectively. For a metal  characterized by a real permittivity, the absorption loss vanishes, and the radiation loss can be easily calculated using the above expressions for $\textsf{T}$ and $\textsf{R}$. For a lossy metal, radiation losses must be calculated independently in order to extract the absorption loss from the expression above. The radiation loss is obtained by integrating the Poynting vector for unit incident flux in region II at $r\rightarrow\infty$:
\begin{equation}\textsf{P}\equiv\int_{\phi_0}^{\theta+\phi_0}{\bf
S}\cdot{\bf\hat{r}}rd\phi.\label{poynting}
\end{equation}
The lower integration limit is set to $\phi_0$ instead of $0$, since the energy radiated from the surface at $\phi=0$ propagates at an angle $\phi_0$ into the far field. Likewise, the upper integration limit is $\theta+\phi_0$ instead of $\theta$. In the short-wavelength limit only the amplitude of the forward-propagating mode is significant, therefore the radiation losses are well approximated by integrating only the counterclockwise propagating mode. A stationary phase approximation is used to obtain an expression for $\phi_0$, using the position-dependent phase $\Phi=k_or+\hbox{Re}[m]\phi$.
%, and the vector normal to a surface of constant phase is obtained by taking the gradient of the phase: ${\bf v}({\bf r})={\bf\nabla}\Phi=k_o{\bf\hat{r}}+\hbox{Re}[m]/r{\bf\hat{\phi}}$, where ${\bf\hat r}$ and ${\bf\hat {\phi}}$ are the unit vector in the radial and angular direction at position ${\bf r}$, respectively.
The change in angle as the wave propagates a radial distance $\delta r$ is $\delta\phi=\hbox{Re}[m]/(k_or^2)\delta r$, giving $\phi_o=\int_R^{\infty}\hbox{Re}[m]/(k_or^2)dr=\hbox{Re}[m]/{k_oR}$.

Calculations were carried out using typical values of silver ($\epsilon_i=-15+i0.5$) in air ($\epsilon_o=1$) with $\omega R/c=800$ and $\theta=90^{\circ}$. Assuming that the metal is lossless ($\epsilon_i=-15$), we find that most of the incident SPP energy is transmitted with $\textsf{T}=0.997$, $\textsf{R}=1.19\times 10^{-8}$, and $\textsf{P}\approx 0.003$. When the absorption loss is accounted for the results change drastically to $\textsf{T}=0.0516$, $\textsf{R}=1.18\times 10^{-6}$, and $\textsf{P}\approx 0.00282$. This result implies that absorption is the dominant loss mechanism when $R\gg\lambda$ even for metals such as silver with relatively low losses. Surprisingly, we find that the overall absorption and radiation losses of SPPs propagating at a non-planar interface may be {\it lower} than the absorption loss of SPPs traveling the equivalent arc distance on a flat surface. This counterintuitive result comes from the fact that the field inside the metal in region II travels an arclength less than $\theta R$ due to the curvature. As a result, SPP fields sample less of the metal volume when propagating on the curved interface than when propagating on a flat surface, resulting in the reduced absorption.

In order to evaluate the accuracy of our results, it is necessary to quantify the validity of the single-mode approximation. As discussed earlier, the single mode approximation is only appropriate when the field profile of the fundamental mode (represented by the solution to the Bessel equation) well approximates the exponentially decaying behavior of the planar SPP. Thus, evaluating the mismatch between the Hankel function and the decaying exponential gives a measure of whether the approximation is appropriate or not. We define the \emph{normalized mismatch} as \begin{equation}\Delta^2\equiv\frac{\int_R^{R+\eta\gamma_o^{-1}}\left|
\exp[-\gamma_o(r-R)]-\frac{H^{(1)}_m(k_or)}{H^{(1)}_m(k_oR)}\right|^2dr}
{\int_R^{R+\eta\gamma_o^{-1}}\left|\exp[-\gamma_o(r-R)]\right|^2dr},
\end{equation}
where $\eta=O(1)$. The expression in the numerator quantifies the field mismatch near the interface. The condition $\Delta\ll 1$ constitutes a criterion for the validity of our approximation. For example, when $\eta=3$, $\epsilon_i=-15$, and $\epsilon_o=1$, $\Delta^2=0.002$ for $\omega R/c=800$, rendering our result applicable. On the other hand, for $\omega R/c=100$ we obtain $\Delta^2=0.3$, indicating that the approximation is less reliable now, hence the coupling to non-fundamental modes can no longer be neglected.

   \begin{figure}[t]
   \includegraphics[width=6cm]{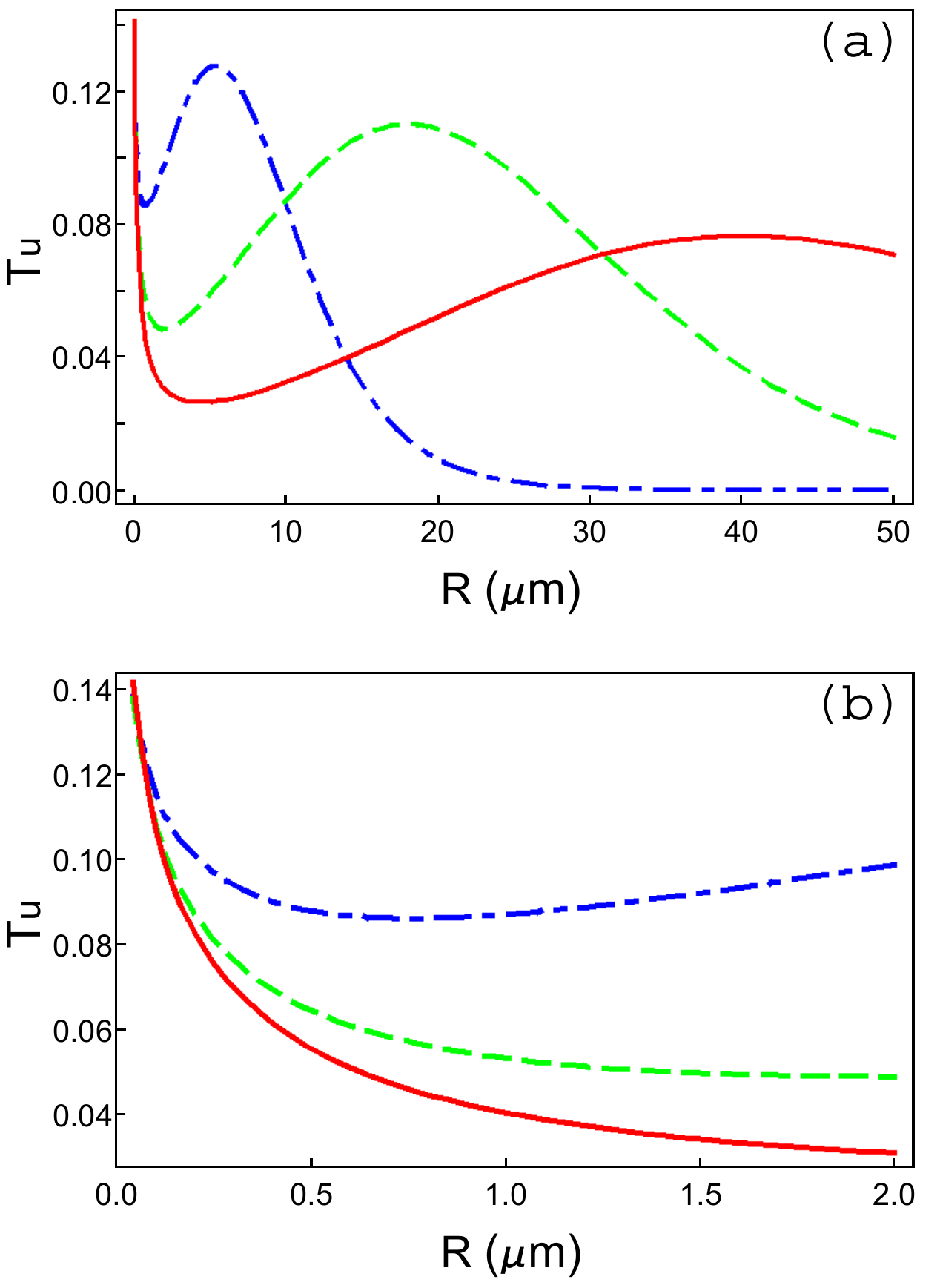}
   \caption{ \label{Fig5}
(a) The upper bound for the transmittance, $\textsf{T}_u$, plotted
for a silver-–air interface with bend angle $\theta=90^{\circ}$, as function
of bend radius $R$ for wavelengths $\lambda=500\hbox{nm}$
(dashed-dotted line), $\lambda=600\hbox{nm}$ (dashed line), and
$\lambda=700\hbox{nm}$ (solid line). (b) Magnified view of the
upper bound in the diffraction-dominated regime.}
   \end{figure}

When the single-mode approximation is not appropriate, it is still possible to derive a physical quantity from the above analysis. As discussed previously, the mode index $m$ associated with the fundamental mode has the smallest imaginary part compared to the mode indices of non-fundamental modes. Since the wave depends on $n$ as $\exp[\pm in\phi]$, modes with large $\hbox{Im}[n]$ decay rapidly. Thus, the transmission in the presence of coupling to nonfundamental modes does not exceed the upper bound of
\begin{equation}
\textsf{T}_u=\exp(-2\hbox{Im}[m]\theta).
\end{equation}
Here we neglect reflections at the $x_1$ and $x_2$ boundaries, thus excluding interference effects. Figure~\ref{Fig5}(a) is a plot of $\textsf{T}_u$ as function of $R$. A peak is clearly visible, moving to higher values of $R$ as the wavelength increases. To the right of the peak, at large radii of curvature absorption losses in the metal dominate, and the maximum transmittance decreases with increasing radius. To the left of the peak radiation due to the high curvature is the dominant loss mechanism, leading to a rapid drop in $\textsf{T}_u$. At very high curvature ($R\leq 10\mu \hbox{m}$) there is a change in trend, and $\textsf{T}_u$ starts to {\em increase} with decreasing $R$. When calculating the radiation loss per arclength, we find that for this range of radii it increases slower than elsewhere, allowing $\textsf{T}_u$ to increase even as $R$ attains very small values.

The behavior of the transmittance for very high curvature is a consequence of diffraction of the incident SPPs. In general, diffraction is associated with the finite wavelength of light, and the diffraction coefficient approaches zero in the limit of small wavelength\cite{Keller62}. For SPPs propagating around bends, a decrease in the radius of curvature is equivalent to an increase in the effective wavelength, $\lambda/R$. For this reason, SPP diffraction around a corner increases as $R$ decreases, resulting in greater transmittance for smaller $R$. Since the diffraction is only significant at sharp corners, the transmittance becomes weakly dependent on the dispersion of the metal at small radii of curvature. Hence, as shown in Fig.~\ref{Fig5}(b), the upper bound at different wavelengths converges to a single value as $R$ approaches zero.

From the discussion above, it is clear that the nonmonotonic behavior of the transmittance is a result of three competing mechanisms: absorption, radiation, and diffraction. Depending on the radius of curvature, one of the three mechanisms becomes dominant, creating three distinct regimes in Fig.~\ref{Fig5}(a). However, it is not possible to evaluate the potential discrepancies between the upper bound model plotted here and the analytical results, since our analytical approach is not valid at high curvatures. To examine the transmittance for such cases, we turn to a numerical study using the finite-difference time-domain method (FDTD).

\section{\normalsize Comparison between analytical results and FDTD calculations}
\label{sec:numerical}

We apply here numerical FDTD calculations to study the propagation of SPPs about bends, and compare with our analytical results. Previously, the method of lines has been used to study the diffraction and the propagation of SPPs at a sharp bend with $R=0$\cite{Jamid95}. Our present numerical study examines how the transmission efficiency depends on the radius of curvature of the bend. The dielectric function in our simulations is given by the Debye model
\begin{equation} \epsilon(\omega)=\epsilon_{\infty}+\frac{\epsilon_s-
\epsilon_{\infty}}{1-i\omega\tau}+i\frac{4\pi\sigma}{\omega}.
\end{equation}
With the choice of parameters $\epsilon_{\infty}=3.90838$, $\epsilon_s=-25658.4$, $\tau=1.20973\times10^{-14}\hbox{sec}$, and $\sigma=1.68931\times 10^{17}\hbox{sec}^{-1}$ the Debye model closely matches experimentally obtained values for silver\cite{Palik85} in the wavelength range 400--1200nm. We implement a nonuniform orthogonal grid, with mesh size in the range $\lambda/300$--$\lambda/40$. Dispersive perfectly matched layers\cite{Fan00} are used as absorbing boundary conditions throughout our study. The wavelength in vacuum is fixed at $\lambda=630\hbox{nm}$. A trial simulation of SPP propagation at a planar metal--air interface indicates that the simulated SPPs are well characterized by the effective dielectric constant $\epsilon_s= -14.0+i0.9$, which is reasonably close to $\epsilon = -15.7 + i1.1$ calculated using the Debye model.

   \begin{figure}[t]
   \includegraphics[width=8.3cm]{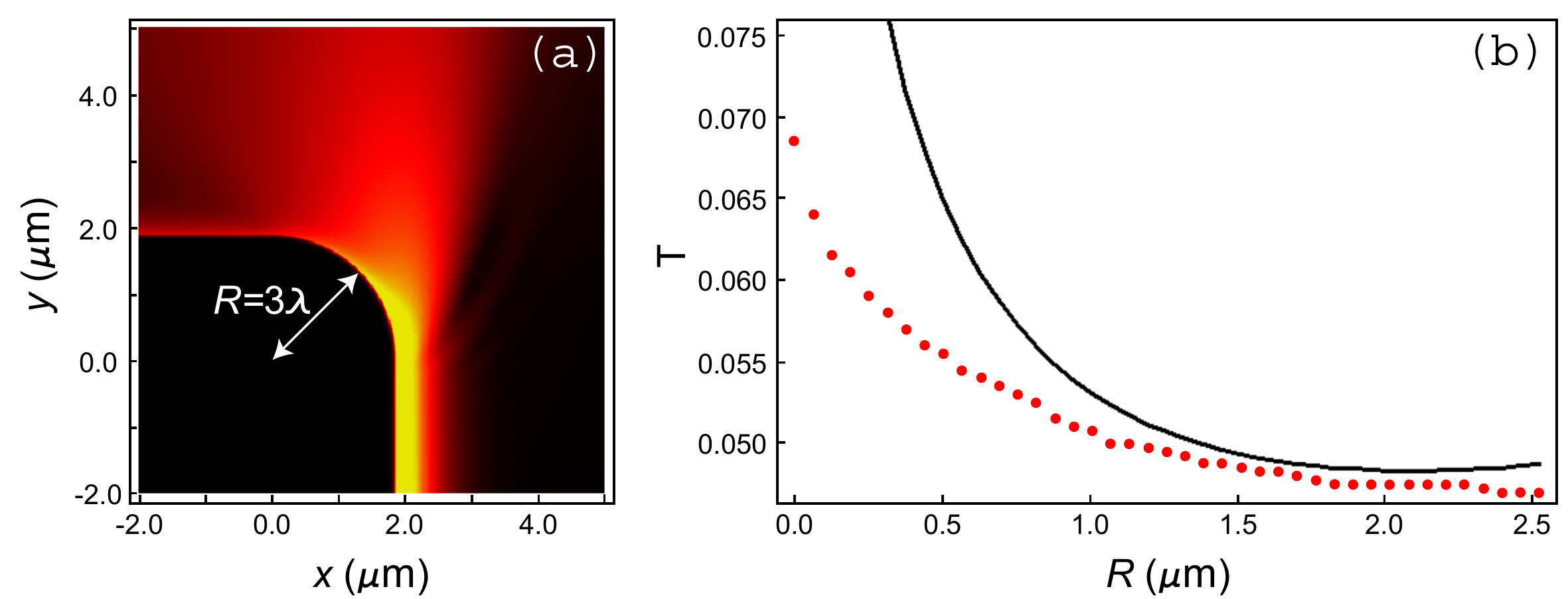}
   \caption{ \label{Fig6} (a) FDTD simulation results showing the magnitude of the magnetic field for $R=\lambda=630\hbox{nm}$. The SPPs are incident from the bottom and propagate counterclockwise around the bend. (b) Analytical result for the transmittance upper bound (black trace) compared to numerical simulations for various values of $R$ (red points). The upper bound values are clearly higher than the simulation results, confirming the consistency of our analytical method.}
   \end{figure}

\subsection{Ninety-degree bend}

A series of FDTD simulations were performed to analyze propagation about the $90^{\circ}$ rounded edge shown in Fig.~\ref{Fig6}(a). The transmittance, $\textsf{T}$, is extracted for various radii of curvature. As shown in Fig.~\ref{Fig6}(b), the transmittance {\it increases} with decreasing radius, peaking at $R=0$ with $\textsf{T}\approx0.07$. Hence, the numerical approach confirms the diffraction-dominated small-radius behavior of $\textsf{T}$ predicted from our earlier analysis of $\textsf{T}_u$ in Fig.~\ref{Fig5}. Interestingly, our simulations reveal that the actual transmittance is reasonably well described by the upper bound (calculated now using $\epsilon_s$) even for $R\leq\lambda$. Moreover, for $R>\lambda$ the discrepancy between the simulation and the analytical upper bound is less than 0.01. This mismatch is expected to decrease even further with decreasing curvature because the coupling to the fundamental mode increases as $\lambda/R\rightarrow 0$. We therefore conclude that the calculated upper bound is a good estimate for the transmittance for all $R\geq\lambda$.

\subsection{Bend with negative curvature}

The analytical formalism developed above may also be used to analyze the {\em reversed geometry}, where the metal occupies the outer space, and the SPPs propagate around a dielectric void in it as shown in Fig.~\ref{Fig1}(c). In this complementary picture the planar SPP modes are now matched to the solutions of a {\em hollow cylindrical void} in the metal. However, care must be taken when choosing the appropriate solutions in region II for the mode matching. For a dielectric cylinder surrounded by metal, the radial solutions in the dielectric are the Bessel functions. These functions, except for the one confined to the interface, have field nodes. Since the existence of nodes implies that photons are exchanged between opposing points on the cylindrical interface, these solutions are unphysical in an open geometry such as our system. Therefore, the only solution represented by the Bessel function in region II is the surface confined mode, which lacks any nodes.

As before, the surface confined mode corresponds to the fundamental (SPP) mode, which minimizes the field mismatch with the incident planar SPP under the single-mode approximation. Since the fundamental mode is represented by the Bessel function, Eq. (\ref{findn}) remains unchanged and the expressions for the transmittance and the reflectance are valid for this reversed geometry. Our calculations have shown that this problem is now appropriately analogous to the finite potential barrier model. We also find that the absorption loss in region II is now greater than the absorption of planar SPPs. From the argument above, it also follows that in the single-mode approximation SPP propagation around a bend is {\em nonradiative}. For this reason, SPP transmission through the reversed geometry becomes highly efficient when material losses and reflections at the boundaries are negligible.

\bigskip
   \begin{figure}[t]
   \includegraphics[width=8.3cm]{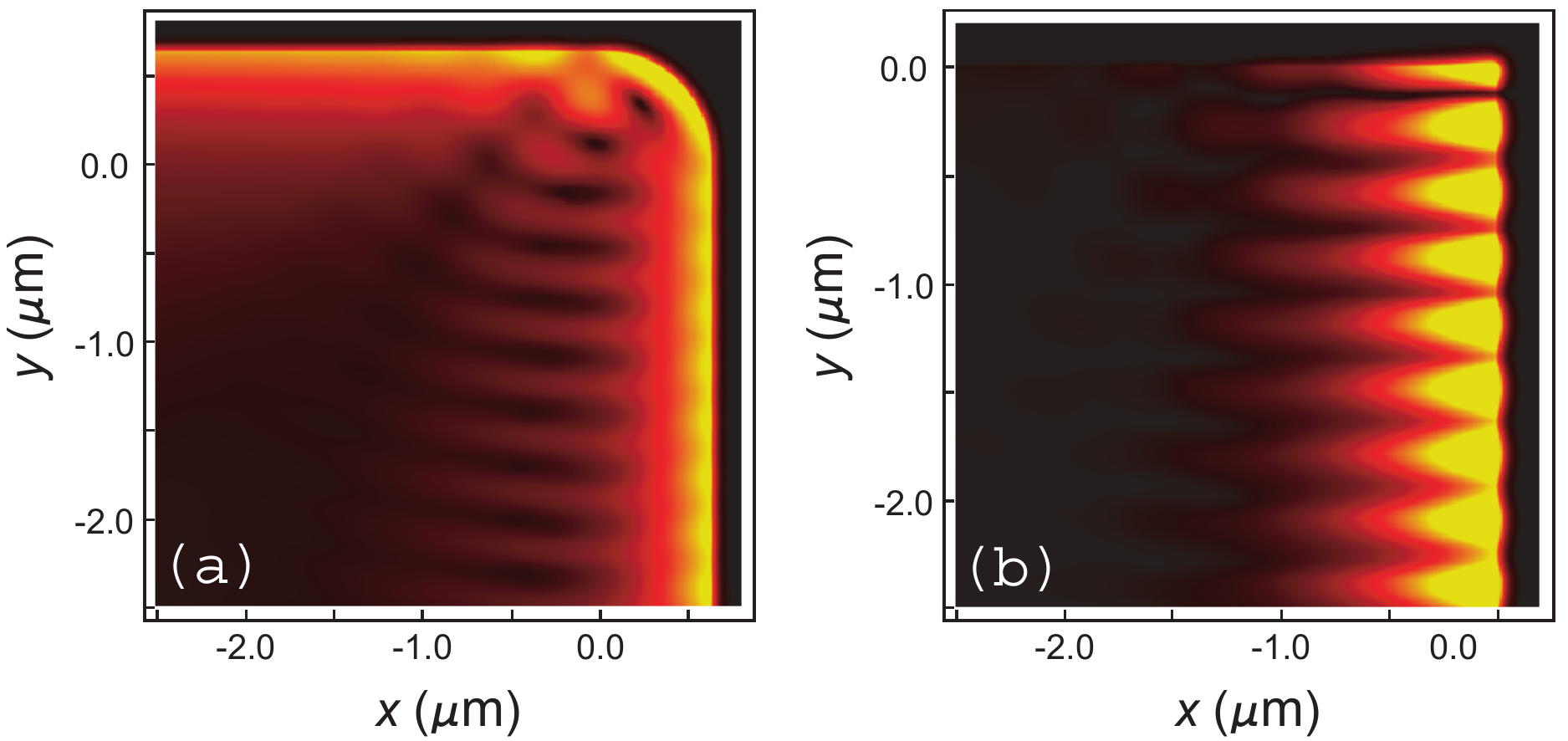}
   \caption{ \label{Fig7}
FDTD simulation results showing the magnitude of the magnetic
field for (a) $R=\lambda=630\hbox{nm}$ and (b) $R=0$ when SPPs, incident
from the bottom, propagate around a circular dielectric void. The
radiation loss is significantly reduced compared to the
propagation around a metallic bend.
    }
   \end{figure}

Our simulations show that efficient propagation is indeed possible. For example, we find that when $R=\lambda$ the transmittance is $\textsf{T}=0.73$, in sharp contrast to the value of $\textsf{T}\approx 0.06$ obtained for an equivalent curvature in the geometry of Fig.~\ref{Fig6}(a). However, when $R=0$, back-reflection at the bend becomes the dominant loss mechanism with $\textsf{R}=0.96$, while the radiation loss remains small as shown in Fig.~\ref{Fig7}(b). The transmittance decreases to $\textsf{T}\approx 0.002$. Because of its high reflectance, this particular geometry is essentially a SPP mirror and may be implemented to construct SPP resonators\cite{Schroter97}.

   \begin{figure}[b]
   \includegraphics[width=8cm]{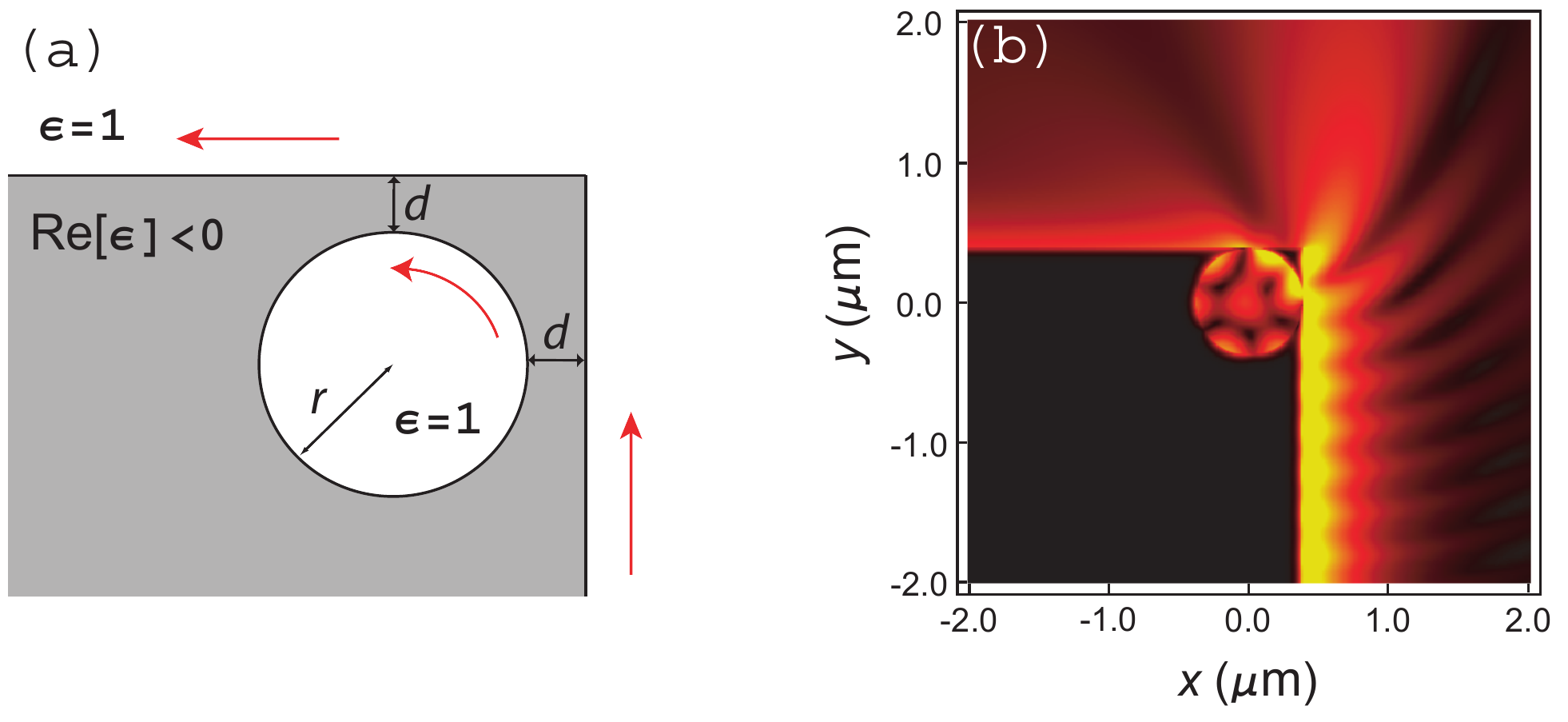}
   \caption{ \label{Fig8}
(a) Schematic diagram of a SPP microresonator embedded near a sharp
edge. SPPs are incident from the bottom along the direction of the
red arrow, couple to the resonator and couple out again. (b)
Numerical FDTD simulation showing the magnitude of the magnetic
field for $d=8.4nm$ and $r=378nm$. A significant portion of the field is seen to couple
into the resonator.
    }
   \end{figure}

\section{\normalsize Resonator-enhanced transmission around a sharp bend}
\label{sec:resonator}

We have shown that efficient transmission of SPP energy around bends is possible when propagating around small voids in metals, while SPP propagation about metallic bends is highly radiative. In what follows we demonstrate a method for reducing radiation losses by introducing an additional metal-dielectric interface into the system, which provides an alternative, low-loss transmission channel for SPPs. Consider the geometry shown in Fig.~\ref{Fig8}(a). A cylindrical hole of radius $r$ is placed in close proximity to two flat metal interfaces joined by an abrupt $90^{\circ}$ edge. Incident SPPs propagating upward along the vertical interface are coupled through their near-fields into the void and our of it onto the second interface. Positioning of the resonator within the skin depth of the metal allows efficient excitation of the resonator modes by the incident SPPs, and subsequent outcoupling. The role of the dielectric void is similar to that of a dielectric microring resonator, and the evanescent wave coupling scheme is analogous to optical coupling between microring resonators and dielectric waveguides\cite{Yariv00}. The SPP resonator provides an alternative transmission channel into which the field couples, leading to a reduction in the total SPP energy impinging on the highly radiative sharp edge. Hence, the efficiency of propagation increases significantly when a resonator is properly incorporated into the metal in the vicinity of the bend. Figure~\ref{Fig8}(b) shows the results of a simulation for a typical resonator-coupled system. We find that the transmission around an infinitely sharp bend in the absence of a resonator, ($\textsf{T}\approx 0.07$) increases to $\textsf{T}=0.17$ when a cavity of radius $r=378nm$ is incorporated at a distance $d=8.4nm$ from the interfaces. A scattering-theory formalism has been previously developed to treat waveguide-resonator couplings\cite{Fan99,Xu00}. However, this method requires ab-initio knowledge of the scattering-matrix elements, attainable by solving Maxwell's equations using FDTD or other numerical methods. When applied to the geometries of our system, we find that this approach is less straightforward than our demonstrated method of extracting the transmission efficiency directly from FDTD simulations of resonator-coupled interfaces. The transmitted signal levels may be then optimized by fine-tuning additional parameters of the system, such as resonator shape, the number of (cascaded) resonators and their relative positioning near the bend. In addition, incorporating a gain medium\cite{Plotz79,Bergman03} (i.e. SPP \emph{amplifier}) inside the resonator cavity may further enhance transmission efficiencies. Future work will address the enhanced transmission via SPP resonators in detail.

\section{\normalsize Conclusions}
\label{sec:conclusions}

In summary, we have studied the propagation of SPPs at a curved metal-dielectric interface both analytically and numerically. We have presented and confirmed a physically intuitive picture of the bend-induced radiation, and how the radiation loss is related to the SPP confinement at the interface. In the short wavelength limit, we have shown that calculating the propagation efficiency is analogous to a quantum mechanical 1D finite square well which is readily solved to obtain the expressions for energy transmission and reflection coefficients. The analysis of the upper bound on the transmittance has revealed that absorption, radiation, and diffraction are three competing mechanisms, responsible for the non-monotonic behavior of the transmittance. Furthermore, our numerical study has shown that the calculated upper bound $\textsf{T}_u$ is a good estimate for the actual transmittance for all bend radii $R\geq\lambda$. In the complementary reversed geometry, when the SPPs propagate around a dielectric void, it is argued that the fundamental mode is nonradiative, and thus SPP propagation is overall less radiative. This prediction was verified numerically. Finally, we have studied a bend geometry in which a surface plasmon resonator is introduced to provide an alternative transmission channel for enhancing the propagation efficiency. Future studies will further analyze resonator design, as well as the details of SPP coupling to resonators in the presence of loss-compensation.

\section*{Acknowledgements}

This work was supported by NSF CAREER Grants ECS-02-39332 and DMR-02-39273, and ARO Grant DAAD19-02-1-0286. K.H. and M.D. wish to thank D. Grischkowsky for stimulating discussions. K.H. also thanks R. Zia for helpful comments.


\begin{thebibliography}{27}
\bibitem[{Foo()}]{Footnote1}
\bibinfo{note}{If the metal is highly absorptive modes with stronger field
  confinement may have larger $\hbox{Im}[n]$, since they experience greater
  absorption loss.}
\end{thebibliography}
\end{document}